\documentclass[
  reprint,                   
  aps,prb,                   
  superscriptaddress,        
  amsmath,amssymb,           
  floatfix, nofootinbib                   
]{revtex4-2}

\usepackage{graphicx}
\usepackage{mathtools}
\usepackage{bm}
\usepackage{braket}
\usepackage{microtype}
\usepackage[acronym]{glossaries}
\usepackage[hidelinks]{hyperref}
\usepackage{cleveref}
\usepackage{amsfonts}
\usepackage{xcolor}
\usepackage{physics}
\usepackage{times}
\usepackage[section]{placeins}
\hypersetup{colorlinks, allcolors=blue}

\newcommand{\ia}{i\alpha}
\newcommand{\jb}{j\beta}

\newcommand{\trans}{%
  ^{\raisebox{0.5ex}{\(\scriptstyle  T\)}}
}

\newcommand{\shortbar}[1]{%
  \mkern1mu\overline{\mkern-1mu #1 \mkern-1mu}\mkern1mu}

\usepackage{etoolbox} 
\pretocmd{\eqref}{Eq.~}{}{} 

\begin{document}

\title{Phonon-mediated spin-polarized superconductivity in altermagnets}

\author{Kristoffer Leraand}
\affiliation{Center for Quantum Spintronics, Department of Physics, Norwegian University of Science and Technology, NO-7491 Trondheim, Norway}

\author{Kristian M{\ae}land}
\affiliation{Institute for Theoretical Physics and Astrophysics, University of W{\"u}rzburg, D-97074 W{\"u}rzburg, Germany}

\author{Asle Sudb{\o}}
\email{asle.sudbo@ntnu.no}
\affiliation{Center for Quantum Spintronics, Department of Physics, Norwegian University of Science and Technology, NO-7491 Trondheim, Norway}

\begin{abstract}
We consider the possibility of phonon-mediated unconventional superconductivity in a recently discovered new class of antiferromagnets, dubbed altermagnets. Within a weak-coupling approach, and using a minimal Lieb lattice model for altermagnets, we find a dominant superconducting instability odd in momentum and even in spin with spin-polarized Cooper pairs. We discuss the origin of this unusual result in terms of the spin-structure of the altermagnetic Fermi surface, in combination with the momentum-space structure of the effective phonon-mediated electron-electron interactions on the Fermi surface. 
\end{abstract}
    
\maketitle

\section{Introduction}More than a century after its discovery \cite{onnes1911leiden, vanDelft2010Sep}, superconductivity continues to be one of the most intensely studied topics in condensed matter physics, with deep connections to the most fundamental descriptions of matter \cite{Anderson,Englert-Brout,Higgs,Guralnik-Hagen-Kibble}. This macroscopic quantum phenomenon is characterized by zero electrical resistance and Higgs condensation, rendering the photon massive \cite{Meissner1933Nov,Anderson,Higgs} below some critical temperature $T_c$.  
Superconductors arising out of good metals with small correlation effects
(conventional low-$T_c$ superconductors with a nodeless $s$-wave superconducting gap), are well described by the Bardeen-Cooper-Schrieffer (BCS) theory \cite{BCS}. Within BCS theory the phenomenon arises from an instability of the Fermi surface (FS) due to effective attraction between electrons.
Originally, exchange of phonons mediated the attraction.
The discovery of unconventional superconductivity with nodal gaps in
strongly correlated fermion systems such as heavy fermions \cite{PhysRevLett.43.1892,WHITE2015246} and high-$T_c$ superconductivity \cite{Bednorz-Mueller,Pines1993,SCspinHistory_Scalapino1999, SCAFMspinMoriya2003Jul, SCspinGapSym_Hirschfeld2011} highlighted
that other bosons could also be responsible for the pairing. In unconventional superconductors \cite{Sigrist1991Apr}, the pairing mechanism often involves complex interactions, such as spin fluctuations, electronic correlations, or orbital effects, leading to non-trivial symmetry and momentum-dependent superconducting gaps. In the high-$T_c$ cuprates, nodes in the gap on the FS have been established through phase-sensitive measurements
\cite{vanHarlingen}, 
establishing that the gap is spin-singlet with $d_{x^2-y^2}$-wave symmetry. 
Furthermore, low-$T_c$, possibly $p$-wave, spin-triplet pairing in itinerant ferromagnets has been predicted and observed \cite{ginzburg1957FMSC, Fay1980FMSC, Jian2009FMSC, Saxena2000FMSCexp, Huy2007FMSCexp}.
Finally, magnon-mediated unconventional superconductivity in heterostructures of magnetic insulators and various gapless fermionic systems has been considered extensively \cite{KargarianFMTI, Hugdal2018May, ArneFMNM, Fjaerbu2019arneAFMNM, EirikAFMNM, ThingstadEliashberg, Sun2023Aug, Sun2024Mar, Hugdal2020FMTI,  Erlandsen2020TIAFM, brekke2023interfacial, MS23, Maeland2023Dec, VinasBostrom2024May, Thingstad2024May}.

Recently, altermagnets have emerged as a new class of magnetic materials that defy the traditional dichotomy of collinear ferromagnetism and antiferromagnetism \cite{Smejkal-Sinova-Jungwirth,BBS23,Agterberg24, osumi2023MnTe, LeePRL2024MnTe, krempasky2024MnTe, reimers2023CrSb, Ding2024CrSb, FZhang, Jiang2024AMLiebExp, Wei2024AMLiebExp, Regmi2025CoNb4Se8, Dale2024CoNb4Se8ARPES}. Altermagnets exhibit time-reversal symmetry breaking and spin-split band structures, yet maintain compensated magnetic ordering without net magnetization. They combine properties of ferromagnets and antiferromagnets, opening new avenues for superconductivity research \cite{Smejkal2022Dec, mazin2022notes, Zhu2023AMSC, Hong2025, Chakraborty2024FFLOAM, Bose2024AMSC, Zhang2024FFLOAM, BBS23, MBS24}. Importantly, the spin-split electronic bands in altermagnets exhibit symmetries that can support exotic pairing states \cite{Parshukov2025Jul} in the presence of suitable interactions. However, spin-singlet or mixed-spin spin-triplet superconductivity requires finite-momentum Cooper pairs or odd-frequency superconductivity \cite{Linder2019Oddw, Chakraborty2024FFLOAM, Hong2025}. Even-frequency zero-momentum Cooper-pairing requires fully spin-polarized Cooper pairs.

Previously, unconventional magnon-mediated superconductivity has been investigated in a minimal model of altermagnets \cite{BBS23, MBS24} on a Lieb lattice \cite{Lieb1989Mar, Antonenko2025Lieb, Jungwirth2024LiebRev, Slot2017LiebExp, Wu2024LiebExp}. 
This particular model exhibits antiferromagnetism along with $d_{x^2-y^2}$-wave symmetric spin splitting of the electron bands. 
The existence of altermagnetism in Lieb lattices has also been confirmed using Hubbard models \cite{Durrnagel2024AMLieb, Kaushal2024AMLieb}. 
While realizing a 2D electronic Lieb lattice has been a longstanding challenge \cite{Slot2017LiebExp, Wu2024LiebExp}, several proposed quasi-two-dimensional (2D) altermagnets display the Lieb lattice in nearly decoupled layers \cite{Kaushal2024AMLieb, FZhang, Jiang2024AMLiebExp, Ni2010LOMSLieb, Wei2024AMLiebExp}.
Magnon-mediated superconductivity 
with zero-momentum Cooper-pairs
in altermagnets, requires invoking double-magnon processes to obtain superconductivity \cite{BBS23, MBS24}.
In a strong-coupling superconductivity scenario, many-body effects on this superconductivity has been investigated \cite{MBS24} using Eliashberg theory of strong-coupling superconductivity \cite{Eliashberg1960Sep, Eliashberg1961}. In this case, strong-coupling renormalizations leads to substantial suppression of the critical temperature $T_c$ \cite{MBS24}. Were the electron-electron interaction to be mediated by a spinless boson, the necessity of a double-boson mediated interaction would be obviated, and the strong-coupling reduction of $T_c$ possibly avoided.

The potential of phonons to yield unconventional superconductivity has been explored in spin-orbit coupled systems, though $s$-wave spin-singlet pairing seems to be preferred from phonons alone \cite{Brydon2014PhononOddParity}.  
Ref.~\cite{Schnell2006dwavephonon} considers a situation where both phonons and spin fluctuations contribute to $d$-wave pairing. 
More recently, it has been demonstrated that 
by including vertex corrections, phonons may mediate $d$-wave spin-singlet in strong-coupling superconductors ~\cite{Schrodi2021dwavephonon}. 

In this work, we predict the emergence of unconventional {\it fully spin-polarized} superconductivity in altermagnets mediated by electron-phonon interactions. The examples mentioned above of phonon-mediated conventional and unconventional spin-singlet superconductivity are not an option in the spin-split band structure of altermagnets, at least not for zero-momentum Cooper-pairs. 
Spin-polarized supercurrents are attractive for the field of superconducting spintronics \cite{Linder2015Scspintronics}, usually realized in complicated heterostructures relying on proximity induced superconductivity. We propose that spin polarized phonon-mediated superconductivity can arise intrinsically in altermagnets.
Our result originates from the interplay between spin-symmetries of the electron bands in altermagnets, and the momentum dependence of the electron-phonon interaction. This provides a new pathway for realizing unconventional superconductivity via a conventional mechanism, and with a substantial $T_c$.

\section{Phonon model}
\begin{figure}[h]
    \centering
    \includegraphics[width=\linewidth]{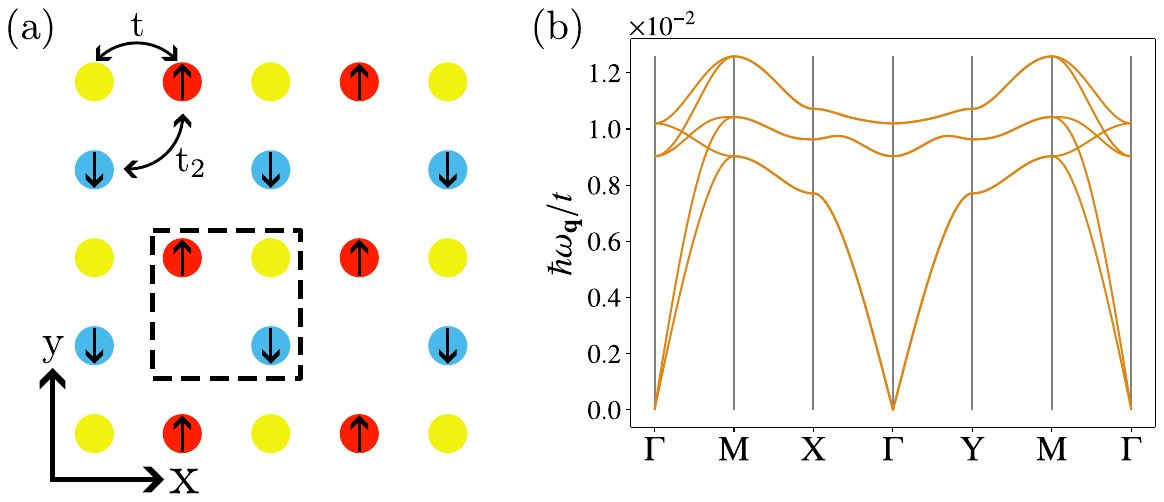}
    \caption{(a) The Lieb lattice, where the black square indicates the chosen unit cell, which contains three atoms with different localized spins, namely, spin-up (red), non-magnetic (yellow), and spin-down (blue). (b) The IP phonon bands for the Lieb lattice on a path between the symmetry points of the Brillouin zone, see Tab.~\ref{tab:Parameters} for the parameter values used.}
    \label{fig:Lieb and Phonon}
\end{figure}

The phonon energies and mode vectors can be found by considering the eigenvalue equation \cite{Lax, BruusFlensberg, Klogetvedt2023, Syljuasen2024}
\begin{equation}
    \label{eq:Eigenvalue of dynamical matrix}
    \sum_\beta D^{\alpha\beta}_\textbf{q}\boldsymbol{\Hat{e}}_{\textbf{q}\beta}^\lambda=M_\alpha\omega_{\textbf{q}\lambda}^2\boldsymbol{\Hat{e}}_{\textbf{q}\alpha}^\lambda,
\end{equation}
where $\alpha,\beta\in\{1,2,3\}$ label the three atoms in the Lieb lattice basis illustrated in Fig.~\hyperref[fig:Lieb and Phonon]{\ref*{fig:Lieb and Phonon}~(a)}, referring to the spin-up, non-magnetic, and spin-down sublattices respectively. $D$ is the dynamical matrix, which is the Fourier transformation of the force constant matrix (FCM). The phonon modes are indicated by $\lambda$, $\boldsymbol{\Hat{e}}_{\textbf{q}}$ is the normalized vector of the Fourier transformed displacement vector, $\textbf{q}$ is the momentum of the phonon, $M_\alpha$ is the mass of atom $\alpha$, and $\omega_{\textbf{q}\lambda}$ is the phonon frequency. 
The symmetries of the Lieb lattice, given in App.~\ref{app:LatSym}, constrain the elements of the FCM \cite{BBS23}, leaving $11$ free parameters. As shown in App.~\ref{app:Phonon properties from symmetries}, by applying additional constraints, we are able to relate these parameters and are only left with one free parameter, $\eta$. This parameter will only affect the phonon bandwidth, which we have chosen to be around $30$~meV. As shown in App.~\ref{app:Electron-phonon}, in our first-order electron-phonon coupling in a 2D system, only the in-plane (IP) phonon modes will couple to the electrons, as such, we discard the out-of-plane (OOP) modes. The IP bands are shown in Fig.~\hyperref[fig:Lieb and Phonon]{\ref*{fig:Lieb and Phonon}~(b)}.   
\section{Electron model}
We utilize a minimal model of a Lieb lattice \cite{Lieb1989Mar, Durrnagel2024AMLieb, Kaushal2024AMLieb} to describe altermagnetic behavior, where itinerant electrons can hop between nearest-neighbor (NN)- and next-nearest-neighbor (NNN) sites, as well as interact with the localized spins on these sites. 
The electronic tight-binding Hamiltonian of the model is given by \cite{BBS23}
\begin{equation}
\label{eq:Hamiltonian electron}
\begin{aligned}
    H_\mathrm{el} &= t\sum_{\langle i,j\rangle \sigma}c_{i\sigma}^\dag c_{j\sigma}+t_2\sum_{\langle\langle i,j\rangle\rangle\sigma}c_{i\sigma}^\dag c_{j\sigma}\\
    &-J_{\mathrm{sd}}\sum_{i\sigma\sigma'}\textbf{S}_i\cdot c_{i\sigma}^\dag\boldsymbol{\sigma}_{\sigma\sigma'}c_{i\sigma'}-\mu\sum_{i\sigma}c_{i\sigma}^\dag c_{i\sigma},
\end{aligned}
\end{equation}
\noindent
where $c_{i\sigma}^{(\dag)}$ annihilates (creates) an electron with spin $\sigma$ at site $i$. The pairs $\langle i,j\rangle$ and $\langle\langle i,j\rangle\rangle$ denote nearest- and next-nearest neighbors with hopping amplitudes $t$ and $t_2$, respectively. $J_{\mathrm{sd}}$ is the on-site exchange coupling between the itinerant electrons and the localized spins $\textbf{S}_i$, and $\boldsymbol{\sigma}$ is the vector consisting of Pauli matrices. The spins $\textbf{S}_i$ have a length $S$, and the on-site energy is set by the chemical potential $\mu$. 
We assume that spin-orbit coupling is negligible compared to the spin-splitting.

\begin{figure}[b]
    \centering
    \includegraphics[width=\linewidth]{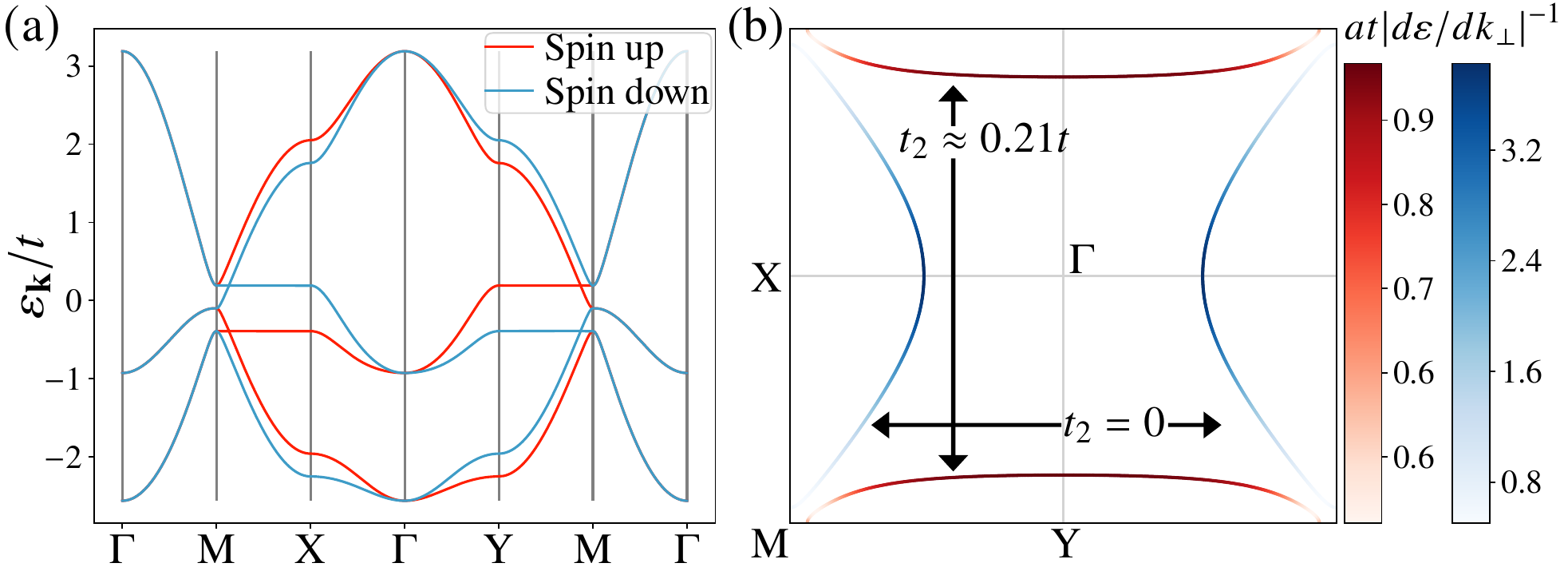}
     \caption{(a) The electronic band structure along a high-symmetry path in the BZ for $\mu=0$. (b) The FSs for spin-up (red) and spin-down (blue), the colors indicate $at\left|d\varepsilon/dk_\perp\right|^{-1}$ for $\mu=0.1t$. The NNN hopping is $t_2\approx0.21t$ for spin-up, and $t_2=0$ for spin-down. This particular system exhibits altermagnetism with $d_{x^2-y^2}$-wave spin splitting. See Tab.~\ref{tab:Parameters} for the rest of the parameters used.}
    \label{fig:Spectrum derivatives}
\end{figure}

The electronic energy spectrum obtained from this Hamiltonian is showcased in Fig.~\hyperref[fig:Spectrum derivatives]{\ref*{fig:Spectrum derivatives}~(a)}, which exhibits the characteristic $d_{x^2-y^2}$-wave spin-split electron bands of an altermagnet. The key feature when finding the spin-polarized superconductivity is the existence of regions that are completely spin-split for all momenta, which can be observed between the middle and top bands. To obtain this spectrum, we diagonalize the Hamiltonian in momentum space as $H_\mathrm{el}=\sum_{n \textbf{k} \sigma}\varepsilon_{n \textbf{k} \sigma}d^\dag_{n \textbf{k}\sigma}d_{n \textbf{k}\sigma}$, where $\varepsilon_{n \textbf{k} \sigma}$ is the energy of band $n$ for excitations with momentum $\textbf{k}$ and spin $\sigma$, while $d^{\left(\dag\right)}_{n \textbf{k}\sigma}$ are the annihilation (creation) band operators. These operators are related to the electron operators as $d_{n\textbf{k}\sigma}=\sum_\alpha q_{\textbf{k}\sigma\alpha n}^*c_{\textbf{k}\sigma\alpha}$,
where $\alpha$ indicates the different atoms in the basis, and the coefficient $q_{\textbf{k}\sigma\alpha n}$ is the component $\alpha$ in the eigenvector of the Hamiltonian corresponding to band $n$.

Due to the $d_{x^2-y^2}$-wave symmetry of the spin splitting, spin-up and spin-down bands are interchangeable by a rotation of $90^\circ$ in momentum space. 
Hence, all properties of the two spins are connected by this rotation. To showcase the role of NNN hopping, we present the results with a NNN hopping of $t_2\approx0.21t$ for spin-up electrons and the results with $t_2=0$ for spin-down electrons throughout the paper. The electronic band structure for $t_2=0$ can be found in Ref.~\cite{BBS23}. 

\section{Electron-phonon coupling}
\label{sec:Electron-phonon coupling}
The electron-phonon interaction is found by an expansion of the two hopping amplitudes in \eqref{eq:Hamiltonian electron} to first order of displacements away from equilibrium, similar to Ref. \cite{TKWS20}.
\begin{equation}
    \label{eq:Exapnd t}
    t(\textbf{r}_{\ia\jb})\approx t(\boldsymbol{\delta})+\textbf{u}_{ij\alpha\beta}\cdot \boldsymbol{\nabla}_r t(\textbf{r}_{ij\alpha\beta})\Big\vert_{\textbf{r}_{ij\alpha\beta}=\boldsymbol{\delta}},
\end{equation}
where $\textbf{r}_{\ia\jb}$ is the instantaneous separation between atoms $\ia$ and $\jb$, $\boldsymbol{\delta}$ are vectors connecting the atoms in equilibrium, and $\textbf{u}_{ij\alpha\beta}$ is their relative displacement from equilibrium. $\boldsymbol{\nabla}_r$ denotes the gradient with respect to $\textbf{r}_{\ia\jb}$, and $\vert_{\textbf{r}_{ij\alpha\beta}=\boldsymbol{\delta}}$ indicates to evaluate the atom positions in equilibrium. See App.~\ref{app:Electron-phonon} for the full derivation of the electron-phonon coupling. The first term of \eqref{eq:Exapnd t} is the equilibrium hopping used in \eqref{eq:Hamiltonian electron}, while the second term is the source of the electron-phonon interaction. We quantize the displacements as \cite{Lax, BruusFlensberg}
\begin{equation}
    \label{eq:Displacement phonon}
    \textbf{u}_{i\alpha} = \sum_{\textbf{q}\lambda}\sqrt{\frac{\hbar}{2M_\alpha N\omega_{\textbf{q}\lambda}}}\boldsymbol{\Hat{e}}_{\textbf{q}\alpha}^\lambda\left(a_{-\textbf{q},\lambda}^\dag + a_{\textbf{q}\lambda}\right)e^{-i\textbf{q}\cdot \textbf{R}_{i\alpha}},
\end{equation}
where $a_{\textbf{q}\lambda}^{(\dag)}$ annihilates (creates) a phonon with momentum $\textbf{q}$ in mode $\lambda$,
$N$ is the number of unit cells, 
and $\textbf{R}_{\ia}$ is the equilibrium position of atom $\alpha$ in unit cell $i$. 

By substituting \eqref{eq:Exapnd t} in \eqref{eq:Hamiltonian electron}, as well as utilizing the band operators, we find the electron-phonon interaction
\begin{equation}
    \label{eq:El-ph}
    H_\mathrm{el-ph} = \sum_{\textbf{k}\textbf{q} \sigma\lambda}g_{\sigma\lambda}\left(\textbf{k}, \textbf{k}+\textbf{q}\right)\left(a_{-\textbf{q}\lambda}^\dag + a_{\textbf{q}\lambda}\right)d_{\textbf{k}+\textbf{q},\sigma}^\dag d_{\textbf{k}\sigma},
\end{equation}
where the electron-phonon coupling strength is given by
\begin{equation}
    \label{eq:g}
    \begin{aligned}
        &g_{\lambda\sigma}\left(\textbf{k},\textbf{k}+\textbf{q}\right) = g_{\lambda\sigma}^{(1)}\left(\textbf{k},\textbf{k}+\textbf{q}\right)+g_{\lambda\sigma}^{(2)}\left(\textbf{k},\textbf{k}+\textbf{q}\right),\\
        &g_{\lambda\sigma}^{(j)}\left(\textbf{k},\textbf{k}+\textbf{q}\right) =\frac{t_j}{2}\sqrt{\frac{\hbar}{2N\omega_{\textbf{q}\lambda}}}\sum_{\alpha\beta}q_{\textbf{k}+\textbf{q},\sigma\alpha}^*q_{\textbf{k}\sigma\beta}\\
        &\times\sum_{\boldsymbol{\delta}_j}e^{i\textbf{k}\cdot\boldsymbol{\delta}_j}\left(\frac{\delta_{jx}}{\sigma_x^2 },\frac{\delta_{jy}}{\sigma_y^2 }\right)\cdot\left[\frac{\boldsymbol{\Hat{e}}_{\textbf{q}\beta}^\lambda}{\sqrt{M_\beta}}e^{i\textbf{q}\cdot\boldsymbol{\delta}_j}-\frac{\boldsymbol{\Hat{e}}_{\textbf{q}\alpha}^\lambda}{\sqrt{M_\alpha}}\right],
    \end{aligned}
\end{equation}
where $j\in\{1,2\}$ indicates whether it is the NN or the NNN coupling, and $\sigma_x, \sigma_y$ are the standard deviations of the atomic orbitals. Note that we have omitted the band indices, as we consider the scattering to be confined to the FS.

\section{Superconductivity}
\label{sec:Superconductivbity}
Next, we transform the electron-phonon interaction into an effective electron-electron interaction by applying a Schrieffer-Wolff transformation \cite{Bardeen1955Aug, SW66}, see App.~\ref{app:Electron-phonon} for the full derivation,
\begin{equation}
    \label{eq:eff Hamiltonian}
    H_{\mathrm{eff}} = \frac{1}{2}\sum_{\textbf{k}\textbf{k}'\sigma}\shortbar{V}_{\textbf{k}'\textbf{k}\sigma}d^\dag_{\textbf{k}'\sigma}d^\dag_{-\textbf{k}',\sigma}d_{-\textbf{k},\sigma}d_{\textbf{k}\sigma},
\end{equation}
where we have followed Refs.~\cite{MS23, MBS24}, and symmetrized the effective interaction as $2\shortbar{V}_{\textbf{k}'\textbf{k}\sigma}\equiv V_{\textbf{k}'\textbf{k}\sigma}+V_{-\textbf{k}',-\textbf{k}\sigma}-V_{-\textbf{k}',\textbf{k}\sigma}-V_{\textbf{k}',-\textbf{k}\sigma}$, with the effective electron-electron interaction on the FS of spin $\sigma$, $V_{\textbf{k}' \textbf{k} \sigma}$ given by 
\begin{equation}
    \label{eq:V}
    V_{\textbf{k}' \textbf{k} \sigma}=-\sum_{\lambda}\frac{g_{\lambda\sigma}\left(\textbf{k},\textbf{k}'\right)g_{\lambda\sigma}\left(-\textbf{k},-\textbf{k}'\right)}{\hbar\omega_{\textbf{k}'-\textbf{k},\lambda}}.
\end{equation}

\begin{figure}[tb]
    \centering
    \includegraphics[width=\linewidth]{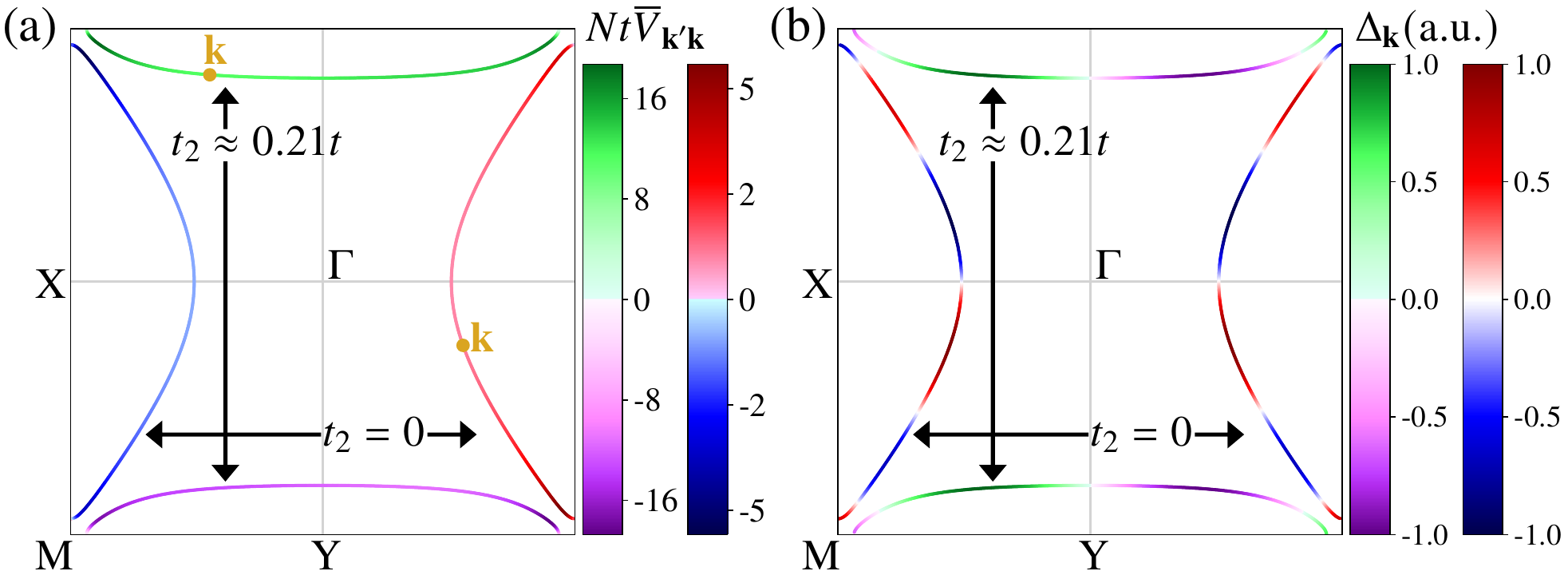}
    
    \caption{(a) shows $N\shortbar{V}_{\textbf{k}'\textbf{k}\sigma}$ for both spin-up (green and purple) and spin-down (red and blue). The values for $\textbf{k}$ are given in yellow and correspond to the maximum of the gap, while the position along the Fermi surface gives $\textbf{k}'$. (b) illustrates the normalized superconducting gap in arbitrary units for both spin-up, and spin-down with the same color coding as in (a). In both plots, the chemical potential is set to $\mu=0.1t$, see Tab.~\ref{tab:Parameters} for the other parameters.}
    \label{fig:V and delta}
\end{figure}

As shown in Refs.~\cite{BBS23, MBS24} the electrons have the greatest DOS for $-J_{\text{sd}}S < \mu < J_{\text{sd}}S$ so we expect the greatest $T_c$ here. In the region $0 < \mu < J_{\text{sd}}S$, there is no overlap of the spin up and spin down FSs, also with $t_2 \neq 0$. Hence, spin-singlet zero-momentum pairing is ruled out, and we focus on this region. The most likely scenario with the type of FS of our model appears to be zero-momentum pairing. This provides the greatest possible phase space for interactions, and, unlike other models, there is a very small overlap of the spin up and spin down FSs even with a finite momentum, App.~\ref{app:FFLO}.
Hence, we look for spin-triplet pairing, which necessitates odd-parity gap functions,
and hence an odd-parity symmetrized coupling \cite{Sigrist1991Apr, MS23}. 
The symmetrized interaction is shown in Fig.~\hyperref[fig:V and delta]{\ref*{fig:V and delta}~(a)} and depicts a $p$-wave symmetry.

We consider weak-coupling superconductivity and apply the linearized BCS equation to solve for the spin-polarized superconducting gap $\Delta_{\textbf{k}\sigma}$ and critical temperature. Since electrons close to the FS dominate the superconducting pairing,
we utilize the FS average of the BCS equation \cite{BBS23, MBS24, Thingstad2021, Maeland2024}
\begin{equation}
\label{eq:BCS FS avgerage}
    \lambda \Delta_{k_\|\sigma} = -\frac{NS_{\mathrm{FS}}}{A_{\mathrm{BZ}}N_\mathrm{samp}}\sum_{k'_\|} \left|\dv{\varepsilon}{k'_\perp}\right|^{-1}\shortbar{V}_{k_\|k'_\|\sigma}\Delta_{k'_\|\sigma},
\end{equation}
where $S_{\mathrm{FS}}$ denotes the length of the FS and $A_{\mathrm{BZ}}$ the area of the first Brillouin zone. $N_{\mathrm{samp}}$ is the number of sampling points, we split $\textbf{k}$ into components $(k_\|,k_\perp)$ tangent and normal to the FS, with $d\varepsilon/dk'_\perp$ being the derivative of the electron energy in the normal direction to the FS.

If we momentarily consider a jellium model for phonons \cite{BruusFlensberg}, the strength of the electron-phonon coupling depends only on the momentum transfer. Still considering spin-polarized superconductivity, the symmetrized coupling would be
$\shortbar{V}_{\textbf{k} \textbf{k}'\sigma} = -|g_{\textbf{k}-\textbf{k}'}|^2/\hbar\omega_{\textbf{k}-\textbf{k}'}+|g_{\textbf{k}+\textbf{k}'}|^2/\hbar\omega_{\textbf{k}+\textbf{k}'}$
on the FS for spin $\sigma$. 

Let us first consider a simple model for acoustic phonons, $g_{\textbf{q}} = g\sqrt{|\textbf{q}|}$ and $\omega_{\textbf{q}} = u |\textbf{q}|$. Then, 
$|g_{\textbf{q}}|^2/\hbar\omega_{\textbf{q}} = g^2/\hbar u$
is momentum independent, and so $\shortbar{V}_{\textbf{k} \textbf{k}' \sigma} = 0$. There is no phonon-mediated spin-polarized Cooper pairing from acoustic phonons within such a simple continuum model. Next, consider a simple model of Einstein phonons, $g_{\textbf{q}} = g|\textbf{q}|$ and $\omega_{\textbf{q}} = \omega_E$. Then,
$\shortbar{V}_{\textbf{k} \textbf{k}' \sigma} = g^2 (|\textbf{k}+\textbf{k}'|^2-|\textbf{k}-\textbf{k}'|^2)/\hbar \omega_{E}.$
Imagining an approximately circular FS around the $\boldsymbol{\Gamma}$ point, the symmetrized interaction closely resembles $\shortbar{V}_{\textbf{k} \textbf{k}' \sigma} =V\cos \phi_{\textbf{k} \textbf{k}'}$, with $\phi_{\textbf{k} \textbf{k}'}$ being the angle between $\textbf{k}$ and $\textbf{k}'$, and $V>0$ some positive energy. 
When solving the eigenvalue problem in \eqref{eq:BCS FS avgerage} for a positive eigenvalue $\lambda$, it is advantageous for the interaction to be attractive when $\textbf{k}'$ is close to $\textbf{k}$, due to the negative sign on the RHS. A coupling of the form $\shortbar{V}_{\textbf{k} \textbf{k}' \sigma} =V\cos \phi_{\textbf{k} \textbf{k}'}$ with $V>0$ prevents non-trivial gap solutions, since the strongest coupling is always at $\textbf{k} = \pm\textbf{k}'$, with a disadvantageous sign.
This illustrates the key difference between phonon- and magnon-mediated interactions, since for magnons typically $|g_{\textbf{k}-\textbf{k}'}| > |g_{\textbf{k}+\textbf{k}'}|$ when $\textbf{k} \approx \textbf{k}'$ due to the strength of the interaction increasing for long-wavelength magnons \cite{EirikAFMNM}. In contrast, for long-wavelength optical phonons, the coupling tends to vanish. 
Hence, in simple continuum models, optical phonons are also unlikely to yield spin-polarized superconductivity. 

The electron-phonon coupling in \eqref{eq:g} depends on both the incoming and outgoing momentum. The simple forms of $g_{\textbf{q}}$ and $\omega_{\textbf{q}}$ of the jellium model are replaced by complex phase factors giving a richer momentum structure on a lattice \footnote{The electron transformation coefficients are real and even in momentum and hence only provide quantitative, not qualitative changes of the symmetrized interaction.}. Thus, both acoustic and optical phonons contribute to the symmetrized pairing, giving a non-zero symmetrized interaction in Fig.~\hyperref[fig:V and delta]{\ref*{fig:V and delta}~(a)}. 

From Fig.~\hyperref[fig:V and delta]{\ref*{fig:V and delta}~(a)}, the interaction has a similar disadvantageous sign as discussed for jellium optical phonons. However, the interaction is now stronger for $\textbf{k} \neq \pm\textbf{k}'$ than at $\textbf{k} = \pm\textbf{k}'$, such as in the corners of Fig.~\hyperref[fig:V and delta]{\ref*{fig:V and delta}~(a)}, which enables solutions for the gap.
If at some $\textbf{k}'\neq\textbf{k}$ we have $\shortbar{V}_{k_\|k'_\|\sigma} > 0$ and $|\shortbar{V}_{k_\|k'_\|\sigma}| > |\shortbar{V}_{k_\|k_\|\sigma}|$, it is possible to have non-trivial solutions of the gap equation if $\Delta_{\textbf{k}\sigma}$ and $\Delta_{\textbf{k}'\sigma}$ have opposite signs, effectively canceling the negative sign on the RHS of \eqref{eq:BCS FS avgerage}. For that reason, our gap solutions tend to have more sign changes on the FS than the effective electron-electron interaction, see Fig.~\ref{fig:V and delta}.

We are interested in the largest eigenvalue of \eqref{eq:BCS FS avgerage}, $\lambda_{\mathrm{m}}$, which is related to the critical temperature by \cite{BCS}
\begin{equation}
    k_B T_c \approx 1.13\hbar\omega_P e^{-\textstyle\frac{1}{\lambda_{\mathrm{m}}}},
\end{equation}
where $\omega_P$ is the largest phonon frequency. Due to the exponential dependence of the critical temperature on the dimensionless coupling, the calculated value of the critical temperature will be sensitive to the chosen values of our parameters. Regardless, our aim is not to pinpoint the exact value of the critical temperature, but rather to show, as a proof of concept, the possibility of achieving finite critical temperatures of phonon-mediated spin-polarized superconductivity.
We parameterize the FS similarly to Ref. \cite{BBS23}, and ensure we use enough sampling points to get convergence of $\lambda_{\mathrm{m}}$ to 3 decimals. 
\begin{figure}[tb]
    \centering
    \includegraphics[width=\linewidth]{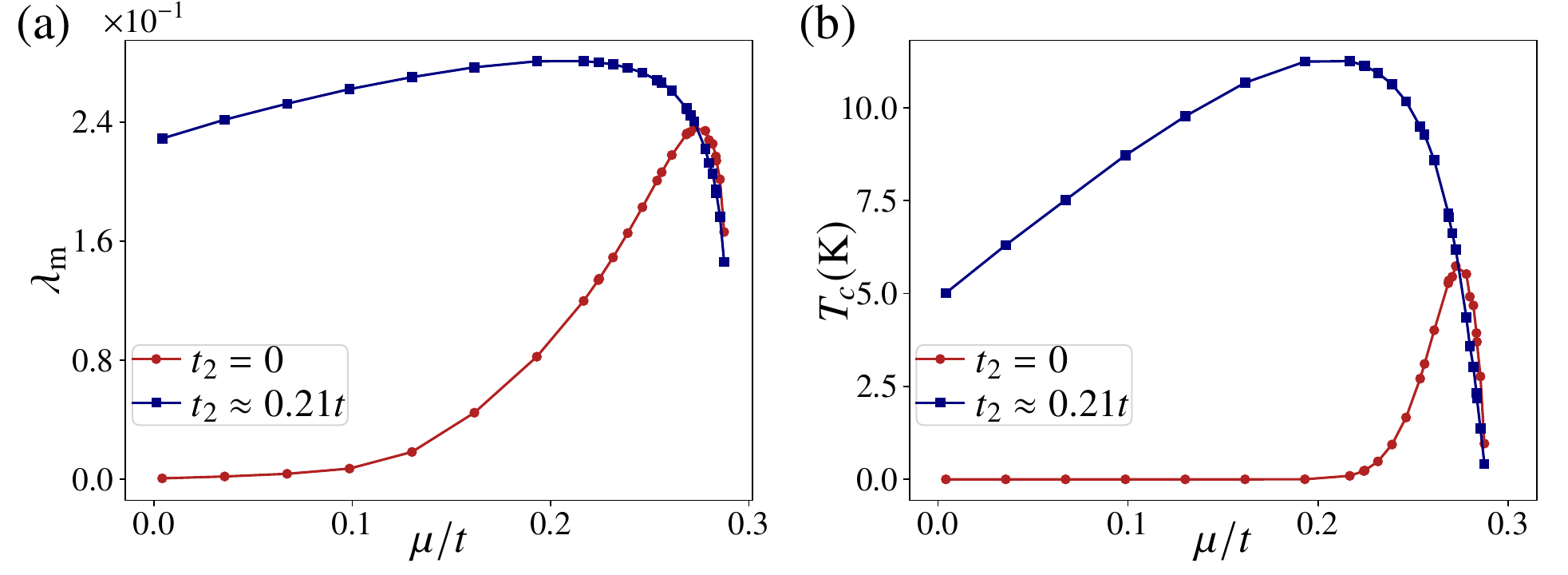}
    \caption{(a) shows the dimensionless coupling and (b) shows the critical temperature, both as functions of the chemical potential for the two cases of including and excluding NNN hopping. See Tab.~\ref{tab:Parameters} for the parameters used.}
    \label{fig:LambdaTc}
\end{figure}
Fig.~\hyperref[fig:LambdaTc]{\ref*{fig:LambdaTc}~(a)} shows that when including the NNN hopping, the dimensionless coupling is much larger for low values of the chemical potential compared to the case $t_2 = 0$. Including NNN hopping means that more electrons couple to the phonons. This can be understood by studying the interactions in Fig.~\hyperref[fig:V and delta]{\ref*{fig:V and delta}~(a)}, where for $t_2=0$, the interaction is much weaker around the center of the arcs and stronger in the corners. The matrix elements of the BCS equation, \eqref{eq:BCS FS avgerage}, are essentially the interaction multiplied by the inverse slope of the band. Comparing the interaction with the inverse slopes in Fig.~\hyperref[fig:Spectrum derivatives]{\ref*{fig:Spectrum derivatives}~(b)}, the inverse slopes are larger at the center of the arcs and weaker towards the edges. However, for $t_2\approx0.21t$, the interaction and the slopes are to a larger degree constant along the arcs, which makes the eigenvalues larger.      

The dimensionless coupling, $\lambda_m$, decreases as the chemical potential approaches the band crossing at 
$\varepsilon=J_{\mathrm{sd}}S$. As shown in Fig.~\hyperref[fig:Spectrum derivatives]{\ref*{fig:Spectrum derivatives}~(b)}, the bands become increasingly flat when approaching the crossing, especially along the lines $\boldsymbol{\Gamma}\textbf{X}$ and $\boldsymbol{\Gamma}\textbf{Y}$, which causes the RHS of \eqref{eq:BCS FS avgerage} to diverge. Combined with the unfavorable sign of the interaction, this makes it increasingly difficult to satisfy the conditions for a gap with a large eigenvalue. 

Careful ARPES measurements of the gap amplitude $|\Delta_{\textbf{k}}|$ would yield information on the nodal structure of the gap. 
Combining this with phase-sensitive measurements can corroborate the presence of nodes \cite{vanHarlingen}.
Additionally, confirming the presence of nodes using STM based methods has been proposed \cite{Sukhachov2023SCandreevSTS, Sukhachov2023SCimpurity}. 
If the number of nodes on the FS per spin corresponds to $p$- or $f$-wave gaps, it would be a strong indication that the mechanism described in this paper is at play.
Note that nodal gaps are sensitive to impurities, and so an experimental realization likely requires a rather clean system.
Identification of nodes can, in principle, be used to distinguish zero-momentum spin-triplet pairing from finite-momentum spin-singlet pairing, since the latter yields $s$-wave or $d$-wave pairing \cite{Chakraborty2024FFLOAM, Hong2025}. 
As noted in Ref.~\cite{Hong2025}, finite momentum pairing is also predicted to yield a distinct Bogoliubov-de Gennes quasiparticle energy spectrum and a distinct density of states compared to zero-momentum pairing. In our case, with a nodal gap, we predict a V-shaped density of states. We emphasize that these predictions should be directly testable on the compounds $\rm{Rb_{1-\delta} V_2 Te_2 O}$, $\rm{K V_2 Se_2 O}$, and ${\mathrm{La}}_{2}{\mathrm{O}}_{3}{\mathrm{Mn}}_{2}{\mathrm{Se}}_{2}$ \cite{Jiang2024AMLiebExp,FZhang, Wei2024AMLiebExp}.

\section{Conclusion}We have shown that phonon-mediated interactions can lead to unconventional, spin-polarized Cooper pairing with substantial $T_c$ in altermagnets. By analyzing the interplay between the unique band structure of altermagnets and electron-phonon coupling, we have found that the distinct spin-splitting in altermagnetic materials enables a novel superconducting state that is fundamentally different from conventional superconductors. The altermagnetic symmetry allows phonons to mediate spin-polarized pairing.
The fully spin-polarized nature of the resulting Cooper pairs represents a significant departure from traditional $s$-wave or spin-singlet superconductivity, offering new possibilities for spintronics and quantum technology applications. Our results further highlight that altermagnetic materials provide a natural platform for exploring unconventional superconducting states, driven by interactions that are traditionally associated with conventional superconductors.

\begin{acknowledgments}
KL and AS were supported by the Research Council of Norway (RCN) through its Centres of Excellence funding scheme, Project No.\ 262633, ``QuSpin'', RCN Project No.\ 323766, as well as COST Action CA21144  ``Superconducting Nanodevices and Quantum Materials for Coherent Manipulation". KM was supported by the DFG (SFB 1170) and the W{\"u}rzburg-Dresden Cluster of Excellence ct.qmat, EXC 2147 (Project-Id 390858490).
\end{acknowledgments}

\appendix

\section{Lattice symmetries} \label{app:LatSym}
We model the altermagnet as a two-dimensional Lieb lattice as proposed in Ref. \cite{BBS23}, a figure of the lattice can be found in Fig.~\hyperref[fig:Lieb and Phonon]{\ref*{fig:Lieb and Phonon}~(a)}.
The Lieb lattice exhibits the eight symmetries
\begin{equation}
\label{eq:Lattice symmetries}
    \begin{aligned}
    \begin{array}{cccc}
        \left(E|0\right), &\left(C_{2z}|0\right), &\left(\sigma_x|0\right), &\left(\sigma_y|0\right),\\[0.5em]
        \left(C_{4z}^+|\mathcal{T}\right), &\left(C_{4z}^-|\mathcal{T}\right), &\left(\sigma_{xy}|\mathcal{T}\right), &\left(\sigma_{\shortbar{xy}}|\mathcal{T}\right),
        \end{array}
    \end{aligned}
\end{equation}
where \textit{E} is the identity operation, $C_{\mathrm{2z}}$ is a twofold rotation, $C_{\mathrm{4z}}^+$ and $C_{\mathrm{4z}}^-$ are fourfold clockwise and anticlockwise rotations respectively, while $\sigma_x$, $\sigma_y$, $\sigma_{xy}$ and $\sigma_{\shortbar{xy}}$ are mirror symmetries about the \textit{x}-axis, \textit{y}-axis, the $y=x$ diagonal, and the $y=-x$ diagonal, respectively. $\mathcal{T}$ is the time-reversal operator, effectively flipping the on-site spins. The notation $\left(A|0\right)$ means that \textit{A} is a symmetry without performing time-reversal, while $\left(B|\mathcal{T}\right)$ indicates that a combination of \textit{B} and time-reversal is a symmetry.

\section{Phonon properties from symmetries}
\label{app:Phonon properties from symmetries} 
The phonon spectrum and eigenmodes are determined in the standard manner 
by studying the displacements $\textbf{u}_{\ia}$ of the lattice sites from their equilibrium positions $\textbf{R}_{\ia}$, where \textit{i} refer to the unit cell and $\alpha$, the different atoms in the basis. We start by expanding the lattice potential around equilibrium, keeping terms up to and including second order in displacements~\mbox{\cite{BruusFlensberg, Syljuasen2024}}
\begin{equation}
\begin{aligned}
    &V(\textbf{r}_{11},\dots, \textbf{r}_{NN_b}) = V(\textbf{R}_{11},\dots, \textbf{R}_{NN_b})\\
    &+\sum_{\ia\mu}\left.\pdv{V}{r_{\ia}^\mu}\right|_{\mathrm{eq}}u^\mu_{\ia}+\frac{1}{2}\sum_{\ia\mu}\sum_{\jb\nu}\left.\frac{\partial^2V}{\partial r_{\ia}^\mu r_{\jb}^\nu}\right|_{\mathrm{eq}}u^\mu_{\ia}u^\nu_{\jb},
\end{aligned}
\end{equation}
where $\textbf{r}_{\ia}=\textbf{R}_{\ia}+\textbf{u}_{\ia}$, is the instantaneous position of the lattice atoms, \textit{N} and $N_b$ denote the number of unit calls and atoms in the basis, respectively. $\mu,\nu\in\{x,y,z\}$ label Cartesian directions, and $|_{\mathrm{eq.}}$ denote evaluation at equilibrium. The first term is a constant, which we can ignore, and the second term is a force, yielding zero when evaluated in equilibrium, so we are only left with the third term. We define the coefficient of the last term as the force constant matrix (FCM) $\Phi^{\alpha\beta}_{\mu\nu}(\textbf{R}_i-\textbf{R}_j)\equiv\left.\partial^2V/\partial r_{\ia}^\mu\partial r_{\jb}^\nu\right|_{\mathrm{eq.}}$.
The FCM is related to the potential of the lattice, hence, it must satisfy the same symmetries as the lattice, namely, \eqref{eq:Lattice symmetries}, which allows us to find constraints on the matrix elements of the FCM. In addition to these symmetries, the FCM satisfies the conditions \cite{Lax, Syljuasen2024}
\begin{align}
    \label{eq:FCM self}
    &\sum_{\jb}\Phi_{\mu\nu}^{\alpha\beta}\left(\textbf{R}_i-\textbf{R}_j\right)=0,\\
    \label{eq:FCM symmetries}
    &\Phi_{\mu\nu}^{\alpha\beta}\left(\textbf{R}_i-\textbf{R}_j\right)=\sum_{\mu'\nu'}S^{{\mathrm T}}_{\mu\mu'}\Phi_{\mu'\nu'}\left[(\mathcal{S}\left(\textbf{R}_{\ia}\right)-\mathcal{S}\left(\textbf{R}_{\jb}\right)\right]S_{\nu'\nu},
\end{align}
where $\mathcal{S}$ is a symmetry of the lattice, so $\mathcal{S}\left(\textbf{R}_{\ia}\right)$ is the transformed vector of $\textbf{R}_{\ia}$ under the symmetry, and $S$ is the matrix representation of this symmetry. We interpret $\Phi^{\alpha\beta}\left(\textbf{R}_i-\textbf{R}_j\right)$ as a bond between atom $\ia$ and atom $\jb$, and separate the symmetries of these bonds into two sets, $G_I$ and $G_R$. $G_I$ is the set of symmetries that leave the bond invariant up to a lattice translation, while $G_R$ contains the symmetries that reverse the bond by switching the positions of the atoms in the bond, up to a lattice translation. The FCM will then satisfy the conditions
\begin{equation}
    \label{eq:FCM symmetry bonds}
    \Phi^{\alpha\beta}\left(\textbf{R}_i-\textbf{R}_j\right)=\left\{
    \begin{array}{ll}
    S\trans\Phi^{\alpha\beta}\left(\textbf{R}_i-\textbf{R}_j\right)S&\quad\text{if}\:S\in G_I,\\
        S\trans\Phi^{\alpha\beta}\left(\textbf{R}_i-\textbf{R}_j\right)\trans S&\quad\text{if}\:S\in G_R.

    \end{array}
    \right.
\end{equation}
Following Ref.~\cite{Syljuasen2024}, we consider connections up to the third nearest neighbors and explicitly find five distinct bonds, while the rest can be expressed in terms of these by applying \eqref{eq:FCM symmetry bonds}.
\begin{widetext}
\begin{align*}
    &\Phi^{11}\left(\left(2a,0\right)\right):\quad 
    & G_I &= \{E, \sigma_y\}, \quad 
    & G_R &= \{C_{2z}, \sigma_x\} \quad 
    & \xRightarrow\quad 
    & \quad\Phi^{11}\left(\left(2a,0\right)\right) = 
    \begin{pmatrix}
        \eta^1_1 & 0 \\
        0 & \eta^1_2
    \end{pmatrix}, \\
    &\Phi^{22}\left(\left(2a,0\right)\right):\quad 
    & G_I &= \{E, \sigma_y\}, \quad 
    & G_R &= \{C_{2z}, \sigma_x\} \quad 
    & \xRightarrow\quad 
    & \quad\Phi^{22}\left(\left(2a,0\right)\right) = 
    \begin{pmatrix}
        \eta^2_1 & 0 \\
        0 & \eta^2_2
    \end{pmatrix},\\
    &\Phi^{33}\left(\left(2a,0\right)\right):\quad 
    & G_I &= \{E, \sigma_y\}, \quad 
    & G_R &= \{C_{2z}, \sigma_x\} \quad 
    & \xRightarrow\quad 
    & \quad\Phi^{33}\left(\left(2a,0\right)\right) = 
    \begin{pmatrix}
        \eta^3_1 & 0 \\
        0 & \eta^3_2
    \end{pmatrix}, \\
    &\Phi^{12}\left(\left(0,0\right)\right):\quad 
    & G_I &= \{E, \sigma_y\}, \quad 
    & G_R &= \varnothing \quad 
    & \xRightarrow\quad 
    & \quad\Phi^{12}\left(\left(0,0\right)\right) = 
    \begin{pmatrix}
        \gamma_1 & 0 \\
        0 & \gamma_2
    \end{pmatrix}, \\
    &\Phi^{13}\left(\left(0,0\right)\right):\quad 
    & G_I &= \{E\}, \quad 
    & G_R &= \{\sigma_{xy}\} \quad 
    & \xRightarrow\quad 
    & \quad\Phi^{13}\left(\left(0,0\right)\right) = 
    \begin{pmatrix}
        \rho_1 & \rho_2 \\
        \rho_3 & \rho_1
    \end{pmatrix}. \\
\end{align*}
\end{widetext}
This method leaves, in principle, eleven free parameters, where $\eta$ is the bond strength between atoms of the same type, $\gamma$ is the NN strength, and $\rho$ is the NNN strength. To narrow down the number of free parameters, we consider the bond strength between all NN, etc., to be equal, hence $\eta_i^j=\eta$, $\gamma_i=\gamma$, and $\rho_i=\rho$. When determining the ratio of $\eta$, $\gamma$, and $\rho$, we assume the bond strength to decrease linearly with distance, such that $\gamma=2\eta$ and $\rho=\sqrt{2}\eta$. With these conditions, we only have one free parameter, $\eta$.
We have chosen $\eta=-1.8\mathrm{N/m}$, so that the phonon bandwidth is about $30$~meV, as illustrated in Fig.~\hyperref[fig:Lieb and Phonon]{\ref*{fig:Lieb and Phonon}(b)}. Also, due to the condition of equal bond strength for the three basis atoms, we have set all their masses equal, $M_\alpha=12~\mathrm{u}\,\forall\,\alpha$.

The rest of the FCM elements can be found by utilizing \eqref{eq:FCM symmetry bonds}, which gives
{\allowdisplaybreaks
\begin{align*}
    &\Phi^{11}\left(\left(0,2a\right)\right) = \sigma_{xy}\trans\Phi^{33}\left(\left(2a,0\right)\right)\sigma_{xy},\\[0.5ex]
    &\Phi^{11}\left(\left(0,-2a\right)\right)=\sigma_{y}\trans\Phi^{11}\left(\left(0,2a\right)\right)\sigma_{y},\\[0.5ex]
    &\Phi^{11}\left(\left(-2a,0\right)\right)=\Phi^{11}\left(\left(2a,0\right)\right)\trans,\\[0.5ex]
    &\Phi^{22}\left(\left(0,2a\right)\right)=C_{4z+}\trans\Phi^{22}\left(\left(2a,0\right)\right)C_{4z+},\\[0.5ex]
    &\Phi^{22}\left(\left(-2a,0\right)\right)=C_{4z+}\trans\Phi^{22}\left(\left(0,2a\right)\right)C_{4z+},\\[0.5ex]
    &\Phi^{22}\left(\left(0,-2a\right)\right)=C_{4z+}\trans\Phi^{22}\left(\left(-2a,0\right)\right)C_{4z+},\\[0.5ex]
    &\Phi^{33}\left(\left(-2a,0\right)\right)=\sigma_x\trans\Phi^{33}\left(\left(2a,0\right)\right)\sigma_x,\\[0.5ex]
    &\Phi^{33}\left(\left(0,2a\right)\right)=\sigma_{xy}\trans\Phi^{11}\left(\left(2a,0\right)\right)\sigma_{xy},\\[0.5ex]
    &\Phi^{33}\left(\left(0,-2a\right)\right)=\Phi^{33}\left(\left(0,2a\right)\right)\trans,\\[0.5ex]
    &\Phi^{12}\left(\left(2a,0\right)\right)=\sigma_{x}\trans\Phi^{12}\left(\left(0,0\right)\right)\sigma_x,\\[0.5ex]
    &\Phi^{23}\left(\left(0,0\right)\right)=C_{4z+}\trans\Phi^{12}\left(\left(0,0\right)\right)\trans C_{4z+},\\[0.5ex]
    &\Phi^{23}\left(\left(0,-2a\right)\right)=C_{4z-}\trans\Phi^{12}\left(\left(0,0\right)\right)\trans C_{4z-},\\[0.5ex]
    &\Phi^{13}\left(\left(2a,0\right)\right)=\sigma_{x}\trans\Phi^{13}\left(\left(0,0\right)\right)\sigma_x,\\[0.5ex]
    &\Phi^{13}\left(\left(0,-2a\right)\right)=\sigma_{y}\trans\Phi^{13}\left(\left(0,0\right)\right)\sigma_y.\\[0.5ex]
    &\Phi^{13}\left(\left(2a,-2a\right)\right)=\sigma_{\shortbar{xy}}\trans\Phi^{13}\left(\left(0,0\right)\right)\trans\sigma_{\shortbar{xy}}.
\end{align*}
}
What remains to be determined are 
the three self-force elements, which are determined by using \eqref{eq:FCM self}, and are given by
\begin{align*}
    \Phi^{11}_{\left(0,0\right)}&=-2\begin{pmatrix}
        \eta^1_1+\eta_2^3+\gamma_1+2\rho_1&0\\
        0&\eta^1_2+\eta^3_1+\gamma_2+2\rho_1
    \end{pmatrix},\\
    \Phi^{22}_{\left(0,0\right)}&=-2\begin{pmatrix}
        \eta^2_1+\eta_2^2+\gamma_1+\gamma_2&0\\
        0&\eta^2_1+\eta^2_2+\gamma_1+\gamma_2
    \end{pmatrix},\\
    \Phi^{33}_{\left(0,0\right)}&=-2\begin{pmatrix}
        \eta^1_2+\eta_1^3+\gamma_2+2\rho_1&0\\
        0&\eta^1_1+\eta^3_2+\gamma_1+2\rho_1
    \end{pmatrix}.
\end{align*}
This model can be extended to include bonds in the \textit{z}-direction by adding the mirror symmetry of the \textit{z}-plane, $\sigma_z$, in all $G_I$ sets, which will leave an additional independent parameter \cite{Syljuasen2024}. However, as we will see in Sec. \ref{app:Electron-phonon}, the electron-phonon interaction with oscillations in the \textit{z}-direction will be of higher order in the displacements, so we neglect these phonon modes. To solve for the phonon modes and frequencies, we turn to momentum-space, where
the dynamical matrix is the Fourier transformation of the FCM, $D^{\alpha\beta}_{\mu\nu}(\textbf{k})\equiv\sum_i\Phi^{\alpha\beta}_{\mu\nu}(\textbf{R}_i-\textbf{R}_j)e^{-i\textbf{k}\cdot(\textbf{R}_{\ia}-\textbf{R}_{\jb})}$, which will then be a $6\cross6$ matrix with elements
{
\allowdisplaybreaks
\begin{align*}
    D(\textbf{k}) &= \begin{pmatrix}
        D^{11}(\textbf{k})&D^{12}(\textbf{k})&D^{13}(\textbf{k})\\
        D^{12}(\textbf{k})^{\raisebox{0.5ex}{\(\scriptstyle\dag\)}}&D^{22}(\textbf{k})&D^{23}(\textbf{k})\\
        D^{13}(\textbf{k})^{\raisebox{0.5ex}{\(\scriptstyle\dag\)}}&D^{23}(\textbf{k})^{\raisebox{0.5ex}{\(\scriptstyle\dag\)}}&D^{33}(\textbf{k})
    \end{pmatrix},\\[0.5ex]
    D^{11}_{xx}(\textbf{k})&=2\left[\eta^1_1\left(\cos(2k_xa)-1\right)\right.\\
    &\left.+\eta^3_2\left(\cos(2k_ya)-1\right)-\gamma_1-2\rho_1\right],\\[0.3ex]
    D^{11}_{yy}(\textbf{k})&=2\left[\eta^1_2\left(\cos(2k_xa)-1\right)\right.\\
    &\left.+\eta^3_1\left(\cos(2k_ya)-1\right)-\gamma_2-2\rho_1\right],\\[0.3ex]
    D^{22}_{xx}(\textbf{k})&=2\left[\eta^2_1\left(\cos(2k_xa)-1\right)\right.\\
    &\left.+\eta^2_2\left(\cos(2k_ya)-1\right)-\gamma_1-\gamma_2\right],\\[0.3ex]
    D^{22}_{yy}(\textbf{k})&=2\left[\eta^2_2\left(\cos(2k_xa)-1\right)\right.\\
    &\left.+\eta^2_1\left(\cos(2k_ya)-1\right)-\gamma_1-\gamma_2\right],\\[0.3ex]
    D^{33}_{xx}(\textbf{k})&=2\left[\eta^3_1\left(\cos(2k_xa)-1\right)\right.\\
    &\left.+\eta^1_2\left(\cos(2k_ya)-1\right)-\gamma_2-2\rho_1\right],\\[0.3ex]
    D^{33}_{yy}(\textbf{k})&=2\left[\eta^3_2\left(\cos(2k_xa)-1\right)\right.\\
    &\left.+\eta^1_1\left(\cos(2k_ya)-1\right)-\gamma_1-2\rho_1\right],\\[0.3ex]
    D^{12}_{xx}(\textbf{k})&=2\gamma_1\cos(k_xa),\\[0.3ex]
    D^{12}_{yy}(\textbf{k})&=2\gamma_2\cos(k_xa),\\[0.3ex]
    D^{13}_{xx}(\textbf{k})&=4\rho_1\cos(k_xa)\cos(k_ya),\\[0.3ex]
    D^{13}_{xy}(\textbf{k})&=4\rho_2\sin(k_xa)\sin(k_ya),\\[0.3ex]
    D^{13}_{yx}(\textbf{k})&=4\rho_3\sin(k_xa)\sin(k_ya),\\[0.3ex]
    D^{13}_{yy}(\textbf{k})&=4\rho_1\cos(k_xa)\cos(k_ya),\\[0.3ex]
    D^{23}_{xx}(\textbf{k})&=2\gamma_2\cos(k_ya),\\[0.3ex]
    D^{23}_{yy}(\textbf{k})&=2\gamma_1\cos(k_ya).
\end{align*}
}
The dynamical matrix is Hermitian, so the elements that are not listed above and cannot be found by the constraint of hermicity are zero. The eigenvalue problem for the dynamical matrix is given by \eqref{eq:Eigenvalue of dynamical matrix}.

\section{Electron-phonon coupling}
\label{app:Electron-phonon}
Similar to Ref. \cite{TKWS20}, we model the electron-phonon interaction by expanding the hopping parameters $t$ and $t_2$ around the equilibrium position of the lattice. We assume the atomic orbitals to be Gaussian,
\begin{equation}
    \label{eq:Wannier}
    \begin{aligned}
    &\phi_{\textbf{r}_{\ia}}(\textbf{r}) = \frac{1}{\pi^{3/4}\sqrt{\sigma_x\sigma_y\sigma_z}}\\
    &\times\exp(-\frac{\left(x-r_{\ia}^x\right)^2}{2\sigma_x^2}-\frac{\left(y-r_{\ia}^y\right)^2}{2\sigma_y^2}-\frac{\left(z-r_{\ia}^z\right)^2}{2\sigma_z^2}),
    \end{aligned}
\end{equation}
where $\textbf{r}_{\ia}$ denote the instantaneous position of atom $\alpha$ in unit cell \textit{i}, $\textbf{r}$ is the spatial coordinate of the wavefunction, $\sigma_x$, $\sigma_y$, and $\sigma_z$ are standard deviations in the three spatial directions. The hopping amplitudes are given by the overlap between two different atomic orbitals modulated by the crystal potential \cite{mahan}
\footnote{We assume the atomic orbitals are eigenstates of the kinetic term and the on-site part of the crystal potential.}
\begin{equation}
    \label{eq:Hopping}
    \begin{aligned}
    t(\textbf{r}_{\ia}- \textbf{r}_{\jb})=\int d^3r\phi_{\textbf{r}_{\ia}}(\textbf{r})A(\textbf{r})\phi_{\textbf{r}_{\jb}}(\textbf{r})&\\
    =A\exp(-\frac{{r_{ij\alpha\beta}^x}^2}{4\sigma_x^2}-\frac{{r_{ij\alpha\beta}^y}^2}{4\sigma_y^2}-\frac{{r_{ij\alpha\beta}^z}^2}{4\sigma_z^2})&,
    \end{aligned}
\end{equation}
where $\textbf{r}_{\ia\jb}\equiv\textbf{r}_{\ia}-\textbf{r}_{\jb}$, and $A(\textbf{r})$ depends on the crystal potential, which we assume to be constant in space. We choose $A$ such that $t=2.4$~eV for $\sigma_x=\sigma_y=\sigma_z=0.4a$. Note that when we consider non-zero $t_2$, it is found from the above equation with the same $A$, giving $t_2 \approx 0.21t$, while we also consider $t_2 = 0$ by artificially setting it to zero.

We Taylor expand the hopping amplitude to first order in displacements from equilibrium, $\textbf{u}$, as $t(\textbf{r}_{ij\alpha\beta})\approx t(\textbf{R}_{ij\alpha\beta})+\textbf{u}_{ij\alpha\beta}\cdot \boldsymbol{\nabla}t$, where we defined the short-hand notation
$\boldsymbol{\nabla}t\equiv\boldsymbol{\nabla}_r t(\textbf{r}_{ij\alpha\beta})\Big\vert_{\textbf{r}_{ij\alpha\beta}=\textbf{R}_{ij\alpha\beta}}$.
The gradient of the hopping is given by
\begin{align}
    \notag
    \boldsymbol{\nabla}_r t(\textbf{r}_{ij\alpha\beta}) = -A&\exp(-\frac{{r_{ij\alpha\beta}^x}^2}{4\sigma_x^2}-\frac{{r_{ij\alpha\beta}^y}^2}{4\sigma_y^2}-\frac{{r_{ij\alpha\beta}^z}^2}{4\sigma_z^2})\\
    \label{eq:Gradient t}
    &\hspace{1.3ex}\times\left(\frac{{r_{ij\alpha\beta}^x}}{2\sigma_x^2}, \frac{{r_{ij\alpha\beta}^y}}{2\sigma_y^2}, \frac{{r_{ij\alpha\beta}^z}}{4\sigma_z^2}\right),
\end{align}
where we note that at equilibrium, the \textit{z}-component is zero. Such that to first order of the displacements, the electrons do not couple to the OOP phonon modes.
We find the electron-phonon interaction by inserting this expansion in the electron Hamiltonian in \eqref{eq:Hamiltonian electron}. The first term in the expansion will simply give the electron hopping term, while the electron-phonon Hamiltonian is then given by
\begin{equation}
    \label{eq:Electron-phonon Hamiltonian basic}
    H_\mathrm{el-ph} = \sum_{\langle i,j\rangle\sigma\alpha\beta}(\textbf{u}_{i\alpha}-\textbf{u}_{j\beta})\cdot\boldsymbol{\nabla}t c_{i\alpha\sigma}^\dag c_{j\beta\sigma},
\end{equation}
which, after inserting the Fourier transformed operators, quantifying the displacements, and utilizing the electron band operators as described in Sec.~\ref{sec:Electron-phonon coupling}, can be written as
\begin{equation}
    \label{eq:Electron-phonon Hamiltonian}
    H_\mathrm{el-ph} = \sum_{\textbf{k} \textbf{q} \sigma\lambda}g_{\sigma\lambda}^{(1)}(\textbf{k}, \textbf{k}+\textbf{q})\left(a_{-\textbf{q},\lambda}^\dag + a_{\textbf{q}\lambda}\right)d_{\textbf{k}+\textbf{q},\sigma}^\dag d_{\textbf{k}\sigma},
\end{equation}
where we have omitted the band index, as we only consider scattering on the Fermi surface (FS). The electron-phonon coupling is given by 
\begin{equation}
    \label{eq:Electron-phonon coupling}
    \begin{aligned}
        g_{\sigma\lambda}^{(1)}(\textbf{k}, \textbf{k}+\textbf{q}) = \frac{t}{2}\sqrt{\frac{\hbar}{2N\omega_{\textbf{q}\lambda}}}\sum_{\alpha\beta}q_{\textbf{k}+\textbf{q},\sigma\alpha}^*q_{\textbf{k}\sigma\beta }&\\
        \times\sum_{\boldsymbol{\delta}}e^{i\textbf{k}\cdot\boldsymbol{\delta}}
        \left(\frac{\delta_x}{\sigma_x^2 },\frac{\delta_y}{\sigma_y^2 }\right)\cdot\left[\frac{\boldsymbol{\Hat{e}}_{\textbf{q}\beta}^\lambda}{\sqrt{M_\beta}}e^{i\textbf{q}\cdot\boldsymbol{\delta}}-\frac{\boldsymbol{\Hat{e}}_{\textbf{q}\alpha}^\lambda}{\sqrt{M_\alpha}}\right]&,
    \end{aligned}
\end{equation}
where $\boldsymbol{\delta}$ are the equilibrium NN vectors. The same procedure can be done for the NNN hopping $t_2$, which gives an electron-phonon coupling $g_{\sigma\lambda}^{(2)}(\textbf{k}, \textbf{k}+\textbf{q})$, that is almost identical to \eqref{eq:Electron-phonon coupling}, only differing with the substitutions $t\rightarrow t_2$ and $\boldsymbol{\delta}\rightarrow\boldsymbol{\delta}_2$, where $\boldsymbol{\delta}_2$ are the vectors between NNNs at equilibrium. The total electron-phonon coupling is then given by $g_{\sigma\lambda}(\textbf{k}, \textbf{k}+\textbf{q})\equiv g_{\sigma\lambda}^{(1)}(\textbf{k}, \textbf{k}+\textbf{q})+g_{\sigma\lambda}^{(2)}(\textbf{k}, \textbf{k}+\textbf{q})$.

\subsection{Effective electron-electron interaction}
The electron-phonon interaction can be written as an effective electron-electron interaction by applying a Schrieffer-Wolff transformation \cite{SW66}. Let the full Hamiltonian be given by $H=H_0+\zeta H_1$, where $H_0=H_\mathrm{el}+H_\mathrm{ph}$, $H_1 = H_\mathrm{el-ph}$, and $\zeta$ is a smallness parameter for the perturbation. We apply a unitary transformation to the full Hamiltonian as
\begin{equation}
    \label{eq:SW}
    \begin{aligned}
        H' &= e^{-\zeta S}He^{\zeta S}\\
       &= H_0+\zeta\left(H_1+\left[H_0,S\right]\right)\\
       &+\zeta^2\left[H_1,S\right]+\frac{1}{2}\zeta^2\left[\left[H_0,S\right],S\right]+\mathcal{O}\left(\zeta^3\right).
    \end{aligned}
\end{equation}
The terms in linear order of $\zeta$ vanish if we choose \textit{S} such that
\begin{equation}
\label{eq:S condition}
    H_1+[H_0, S] =0,
\end{equation}
which leaves the transformed Hamiltonian to be
\begin{equation}
\label{eq:Effective Hamiltonian}
    H' = H_0 +\frac{1}{2}\zeta^2[H_1, S] + \mathcal{O}\left(\zeta^3\right).
\end{equation}
To find the \textit{S} satisfying \eqref{eq:S condition}, we start by looking at $H_0$ and $H_1$,
\begin{align}
    H_0 &= \sum_{\textbf{k}\sigma }\varepsilon_{\textbf{k}\sigma }d_{\textbf{k}\sigma }^\dag d_{\textbf{k}\sigma } + \sum_{\textbf{q}\lambda}~\hbar\omega_{\textbf{q}\lambda}\left(a_{\textbf{q}\lambda}^\dag a_{\textbf{q}\lambda} + \frac{1}{2}\right),\\
    H_1 &= \sum_{\textbf{k}\textbf{q} \sigma\lambda}g_{\lambda\sigma}\left(\textbf{k}, \textbf{k}+\textbf{q}\right)\left(a_{-\textbf{q},\lambda}^\dag + a_{\textbf{q}\lambda}\right)d_{\textbf{k}+\textbf{q},\sigma }^\dag d_{\textbf{k}\sigma }.
\end{align}
From these expressions, we make an ansatz for \textit{S} as
\begin{equation}
\begin{aligned}
    S = &\sum_{\textbf{k}\textbf{q}\sigma\lambda}g_{\lambda\sigma}\left(\textbf{k}, \textbf{k}+\textbf{q}\right)\\
    \times&\left(x_{\textbf{k}\textbf{q}\sigma\lambda} a_{-\textbf{q},\lambda}^\dag + y_{\textbf{k}\textbf{q}\sigma\lambda }a_{\textbf{q}\lambda}\right)d_{\textbf{k}+\textbf{q},\sigma }^\dag d_{\textbf{k}\sigma },
\end{aligned}
\end{equation}
solving for the coefficients $x_{\textbf{k}\textbf{q}\sigma\lambda }$ and $y_{\textbf{k}\textbf{q}\sigma\lambda }$ by applying the condition \eqref{eq:S condition} gives
\begin{align}
    x_{\textbf{k}\textbf{q}\sigma\lambda} &= \frac{1}{\varepsilon_{\textbf{k}\sigma}-\varepsilon_{\textbf{k}+\textbf{q},\sigma}-\hbar\omega_{\textbf{q}\lambda}},\\
    y_{\textbf{k}\textbf{q}\sigma\lambda} &= \frac{1}{\varepsilon_{\textbf{k}\sigma}-\varepsilon_{\textbf{k}+\textbf{q},\sigma}+\hbar\omega_{\textbf{q}\lambda}}.
\end{align}
Note that since we have assumed the scattering to be contained on the FS, the two electron energies in the denominators will cancel.
We can now find the effective electron-electron interaction in \eqref{eq:Effective Hamiltonian}, where we only keep terms corresponding to electron-electron interactions and neglect higher-order phonon terms,
\begin{equation}
\begin{aligned}
    H_\mathrm{eff} = &\frac{1}{2}\sum_{\textbf{k} \textbf{q} \sigma}\sum_{\lambda\textbf{k}' \sigma'}g_{\lambda\sigma}\left(\textbf{k}, \textbf{k}+\textbf{q}\right)g_{\lambda\sigma'}\left(\textbf{k}', \textbf{k}'-\textbf{q}\right)\\
    &\times\left(y_{\textbf{k}\textbf{q}\sigma\lambda}-x_{\textbf{k}\textbf{q}\sigma\lambda}\right)d_{\textbf{k}+\textbf{q},\sigma}^\dag d_{\textbf{k}'-\textbf{q},\sigma'}^\dag d_{\textbf{k}\sigma} d_{\textbf{k}'\sigma'}.
\end{aligned}
\end{equation}
As explained in Sec.~\ref{sec:Superconductivbity}, we consider spin-polarized, zero-momentum Cooper pairs, which set the conditions $\sigma'=\sigma$ and $\textbf{k}'=-\textbf{k}$.
With these assumptions, the interaction takes the form
\begin{equation}
    H_{\mathrm{eff}} = \sum_{\textbf{k}\textbf{k}'\sigma}V_{\textbf{k}'\textbf{k}\sigma}d_{\textbf{k}'\sigma}^\dag d_{-\textbf{k}',\sigma}^\dag d_{-\textbf{k},\sigma}d_{\textbf{k}\sigma},
\end{equation}
where we defined a new summation variable $\textbf{k}'\equiv \textbf{k}+\textbf{q}$, and the interaction strength is given by 
\begin{equation}
\label{eq:Effective interaction}
    V_{\textbf{k}'\textbf{k}\sigma} = -\sum_\lambda \frac{g_{\sigma\lambda}(\textbf{k},\textbf{k}')g_{\sigma\lambda}(-\textbf{k},-\textbf{k}')}{\hbar\omega_{\textbf{k}'-\textbf{k},\lambda}},
\end{equation}
with the momenta on the FS of spin $\sigma$.
We follow the procedure of Ref. \cite{MBS24} and symmetrize the interaction, giving the effective electron-electron Hamiltonian
\begin{equation}
\begin{aligned}
    H_\mathrm{eff} &=\frac{1}{2}\sum_{\textbf{k}\textbf{k}'\sigma}\shortbar{V}_{\textbf{k}'\textbf{k}\sigma}d_{\textbf{k}'\sigma}^\dag d_{-\textbf{k}',\sigma }^\dag d_{-\textbf{k},\sigma } d_{\textbf{k}\sigma},\\
    2\shortbar{V}_{\textbf{k}'\textbf{k}\sigma}&= V_{\textbf{k}'\textbf{k}\sigma}+V_{-\textbf{k}',-\textbf{k},\sigma}-V_{-\textbf{k}',\textbf{k}\sigma}-V_{\textbf{k}',-\textbf{k},\sigma}.
\end{aligned}
\end{equation}

\section{Details of the effective interaction on the Fermi surface} 
\label{sec:EffIntDet}
\begin{figure}[h]
    \centering
    \includegraphics[width=\linewidth]{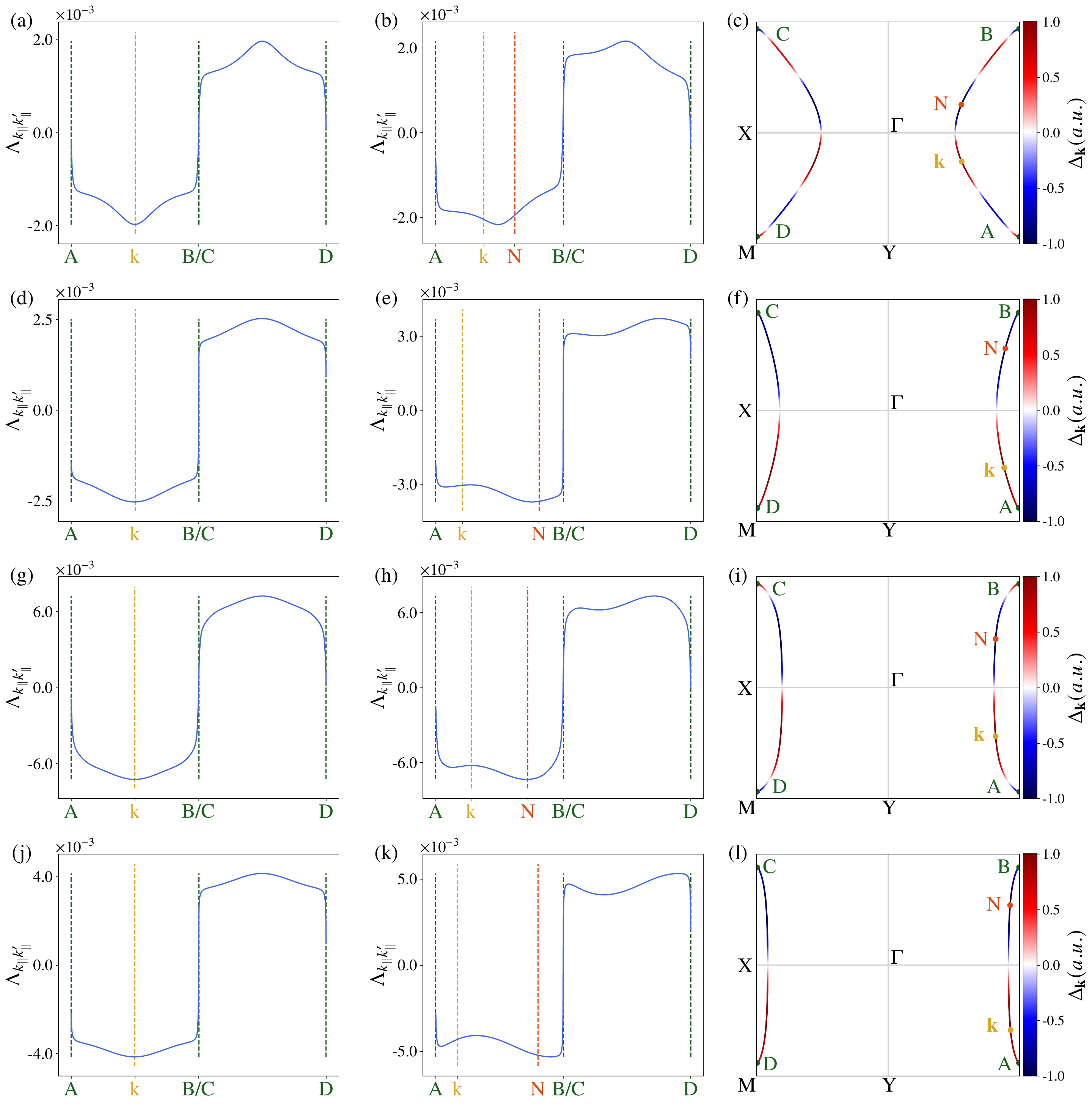}
    \caption{The first column shows the matrix elements $\Lambda_{k_\|k'_\|\sigma}$ for spin-down, with $\textbf{k}$ given by the yellow dashed line, corresponding to the midpoint between \textit{A} and \textit{B} in the respective plot in the third column, and $\textbf{k}'$ is given by the position along the path. The second column has $\textbf{k}$ as marked in the respective third column plots, corresponding to the maximum of the gap. The third column shows the gap for spin-down. The rows are structured such that in the first row (a-c), we have used $\left(t_2=0,\mu=0.1t\right)$, in the second (d-f) $\left(t_2=0,\mu=0.25t\right)$, in the third (g-i) $\left(t_2\approx0.21t,\mu=0.1t\right)$, and in the fourth (j-l) $\left(t_2\approx0.21t,\mu=0.25t\right)$. N indicates the momentum with the most negative value for the gap. Table \ref{tab:Parameters} lists the remaining parameters.}
\label{tableauLambda}
\end{figure}

In this section, we discuss in more detail the momentum dependence of the effective electron-electron interaction on the FS. We introduce the quantity $\Lambda_{k_\|k'_\|\sigma}$ via the linearized Fermi-surface averaged gap equation \cite{BBS23, MBS24, Maeland2024}
\begin{equation}
\begin{aligned}
    \lambda \Delta_{k_\|\sigma} &= -\frac{NS_{\mathrm{FS}}}{A_{\mathrm{BZ}}N_\mathrm{samp}}\sum_{k'_\|} \left|\dv{\varepsilon}{k'_\perp}\right|^{-1}\shortbar{V}_{k_\|k'_\|\sigma}\Delta_{k'_\|\sigma}\\
&\equiv\sum_{k'_\|}\Lambda_{k_\|k'_\|\sigma}\Delta_{k'_\|\sigma}, \label{eq:LambdaGap}
\end{aligned}
\end{equation}
where the meaning of the symbols is explained under \eqref{eq:BCS FS avgerage}. 
This constitutes an eigenvalue problem for a positive quantity $\lambda$ with eigenvector $\Delta_{k_\|\sigma}$, and it is the momentum structure of the matrix $\Lambda_{k_\|k'_\|\sigma}$ that determines whether or not non-trivial solutions exist. Some of these matrix elements are
plotted on the FS in Fig.~\ref{tableauLambda} (left and middle column), along with the eigenvectors $\Delta_{k_\|\sigma}$ on the FS in the right column. Notice how both the symmetrized interaction 
$\shortbar{V}_{\textbf{k}\textbf{k}'\sigma}$
as well as the density of states factor 
$\left|\dv{\varepsilon}{k'_\perp}\right|^{-1}$ 
enters in $\Lambda_{k_\|k'_\|\sigma}$. 
In the first row of Fig.~\ref{tableauLambda} with $t_2=0$ and $\mu=0.1t$, Fig.~\hyperref[fig:LambdaTc]{\ref*{fig:LambdaTc}(a)}, shows that the dimensionless coupling is $\lambda_m \approx 0.01$, which is too low to give a non-negligible $T_c$. From \eqref{eq:LambdaGap} and Fig.~\hyperref[tableauLambda]{\ref*{tableauLambda}~(a)}, we see that $\Lambda_{k_\|k'_\|\downarrow}$ has a disadvantageous sign, as discussed in Sec.~\ref{sec:Superconductivbity}. It is also most substantial for $\textbf{k}' = \pm \textbf{k}$ along the $\boldsymbol{\Gamma}\textbf{X}$ line. Hence, it is natural that the eigenvector with the largest eigenvalue has zero gap on the $\boldsymbol{\Gamma}\textbf{X}$ line. Figure \ref{tableauLambda}(b) shows $\Lambda_{k_\|k'_\|\downarrow}$ when $\textbf{k}$ is placed at the maximal gap amplitude. For that value of $\textbf{k}$, we see that $|\Lambda_{k_\|k'_\|\downarrow}|$ is not maximal at $\textbf{k}' = \pm \textbf{k}$, rather it is maximal for $\textbf{k}'$ on the $\boldsymbol{\Gamma}\textbf{X}$ line. 

By comparing the coupling in Fig.~\hyperref[tableauLambda]{\ref*{tableauLambda}~(b)} with the gap in Fig.~\hyperref[tableauLambda]{\ref*{tableauLambda}~(c)}, we see how appropriate sign changes of the gap generate an eigenvector with a positive eigenvalue $\lambda_m$. 
From Fig.~\hyperref[tableauLambda]{\ref*{tableauLambda}~(b)}, we see that $|\Lambda_{k_\|k'_\|\downarrow}|$ with $\textbf{k}$ placed in the maximum positive gap ($\Delta_{\textbf{k}\downarrow}>0$) and $\textbf{k}'$ placed in the maximum negative gap ($\Delta_{\textbf{k}'\downarrow}<0$, $\textbf{k}' = \mathrm{N}$), is smaller than the maximum $|\Lambda_{k_\|k'_\|\downarrow}|$ in Fig.~\hyperref[tableauLambda]{\ref*{tableauLambda}~(a)}. As a result, the eigenvector is barely able to satisfy the requirement of positive eigenvalue, and $\lambda_m$ is small.

Then, in the second row of Fig.~\ref{tableauLambda} with $t_2=0$ and $\mu=0.25t$, Fig.~\hyperref[fig:LambdaTc]{\ref*{fig:LambdaTc}(a)} shows that $\lambda_m \approx 0.20$, which is larger than the first row by a considerable amount, leading to a non-vanishing critical temperature. This may be understood from the fact that the maximal magnitude of $|\Lambda_{k_\|k'_\|\downarrow}|$ is much greater in Fig.~\hyperref[tableauLambda]{\ref*{tableauLambda}~(e)} than in Fig.~\hyperref[tableauLambda]{\ref*{tableauLambda}~(d)}. Also, unlike the first row [Fig.~\hyperref[tableauLambda]{\ref*{tableauLambda}~(b)}], the peak of $|\Lambda_{k_\|k'_\|\downarrow}|$ in Fig.~\hyperref[tableauLambda]{\ref*{tableauLambda}~(e)} occurs close to where $\textbf{k}$ is placed in the maximum positive gap and $\textbf{k}'$ is placed in the maximum negative gap. Hence, the sign change of the gap function takes significant advantage of the coupling, giving a greater eigenvalue $\lambda_m$.

The situation in the third and fourth rows with $t_2\approx0.21t$, is very similar to the second row. In the third row, however, the difference in magnitudes $|\Lambda_{k_\|k'_\|\downarrow}|$ between column one and two is about the same as in the first row, but an important difference is that the values of $|\Lambda_{k_\|k'_\|\downarrow}|$ are larger, making it possible to have a higher $\lambda_m$. More importantly, the shape of $\Lambda_{k_\|k'_\|\downarrow}$ in Fig.~\hyperref[tableauLambda]{\ref*{tableauLambda}~(h)} is more similar to Figs.\hyperref[tableauLambda]{~\ref*{tableauLambda}~(e)} and \hyperref[tableauLambda]{~\ref*{tableauLambda}~(k)} than Fig.~\hyperref[tableauLambda]{\ref*{tableauLambda}~(b)}. This gives a markedly different momentum structure of the gap in rows two, three, and four [Figs.\hyperref[tableauLambda]{~\ref*{tableauLambda}~(f)}, \hyperref[tableauLambda]{~\ref*{tableauLambda}~(i)}, and \hyperref[tableauLambda]{~\ref*{tableauLambda}~(l)}] compared to row one [Fig.~\hyperref[tableauLambda]{\ref*{tableauLambda}~(c)}]. This difference in the momentum dependence of the coupling and gap function enables a larger eigenvalue $\lambda_m$.

The increased $\lambda_m$ when including NNN hopping is understood from the effect of non-zero $t_2$ on the inverse slope of the FS, shown in Fig.~\hyperref[fig:Spectrum derivatives]{\ref*{fig:Spectrum derivatives}~(b)}. 
Compared to $t_2 = 0$, the slope is not as strongly peaked along the $\boldsymbol{\Gamma}\textbf{X}$ line, reducing the adverse effects of $|\Lambda_{k_\|k'_\|\downarrow}|$ with $\textbf{k}' = \pm \textbf{k}$ on the $\boldsymbol{\Gamma}\textbf{X}$ line. As mentioned in Sec.~\ref{sec:Superconductivbity}, a non-zero $t_2$ also enables more electron-phonon processes, which increases the coupling strength.

\section{Finite-momentum Cooper pairs} \label{app:FFLO}
\begin{figure}[b]
    \centering
    \includegraphics[width=\linewidth, height=0.45\textheight, keepaspectratio]{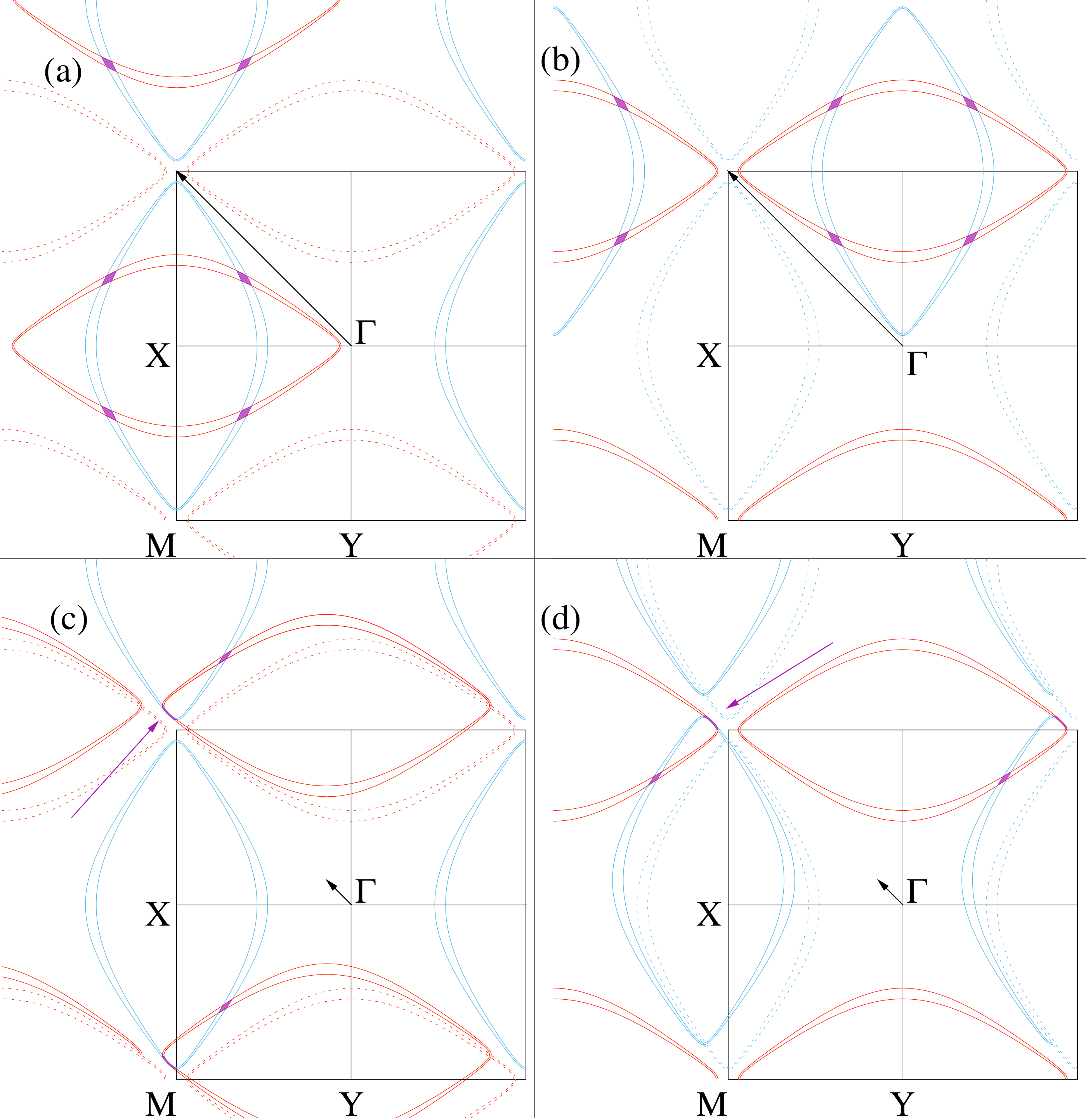}
    \caption{The box marks the BZ, solid lines denote states with energies within a phonon bandwidth of the Fermi level, and the dashed lines indicate which FS is shifted. The overlap between the shifted FS and the opposite spin-FS is marked in purple. The spin-up FS is shifted in (a) and (c), while for (b) and (d), it is spin-down. The vector starting in the $\Gamma$-point is the finite-momentum vector, showing how the FS is shifted. The finite-momentum vector is chosen to maximize the overlap in (a) and (b), while for (c) and (d), it sets the overlap at the ends of the ellipses highlighted by the purple arrow. In all the cases, $\mu=0.1$ and $t_2=0$, as these values give relatively large overlap. See Table \ref{tab:Parameters} for the remaining parameters.}
    \label{Fig:Overlap}
\end{figure}
In the Sec.~\ref{sec:Superconductivbity}, we have focused on spin-polarized pairing. However, there is still a possibility for both singlet and spin-zero triplet states. To form different-spin Cooper pairs, a finite-momentum state is required, effectively shifting one of the spin-FS in Fig.~\hyperref[fig:Spectrum derivatives]{\ref*{fig:Spectrum derivatives}(a)} on top of the other, as illustrated in Fig.~\ref{Fig:Overlap}. Even in the case of maximum overlap, there is a significantly smaller area as opposed to the spin-polarized case, where the whole region between the solid lines can form Cooper pairs. In the cases of Fig.~\hyperref[Fig:Overlap]{\ref*{Fig:Overlap}~(a)} and \hyperref[Fig:Overlap]{(b)}, where we have maximized the overlap, the intersection area is $0.073$ times the area of the spin-polarized case. While for the case of Fig.~\hyperref[Fig:Overlap]{\ref*{Fig:Overlap}~(c)} and \hyperref[Fig:Overlap]{(d)}, the intersection area is $0.016$ times the spin-polarized area.
Let us compare this to Fig.~1(b) in Ref.~\cite{Hong2025}. In their model, the ellipses have a rather large overlap for the case comparable to our Fig.~\hyperref[Fig:Overlap]{\ref*{Fig:Overlap}~(c)} and \hyperref[Fig:Overlap]{(d)}. The Lieb lattice model has much sharper edges of the ellipses here, and much more dispersive bands in this region compared to the rest of the ellipse, both of which significantly reduces the likelihood of finite-momentum pairing compared to the model of Ref.~\cite{Hong2025} and other models that have studied the competition of zero-momentum and finite-momentum pairing in altermagnets \cite{Chakraborty2024FFLOAM}. Also note that the case in Fig.~\hyperref[Fig:Overlap]{\ref*{Fig:Overlap}~(a)} and \hyperref[Fig:Overlap]{(b)} is comparable to Fig.~1(a) in Ref.~\cite{Hong2025} with 4 discrete overlap points of the FSs, only that in our case it also requires finite momentum since the spin up and spin down FSs have no overlap originally. Such a state is not expected to be competitive at strong spin splitting \cite{Hong2025}, which is the case we focus on.

\section{Numerical parameters}
\label{sec:Params}
Table \ref{tab:Parameters} shows the value of parameters that are the same throughout the paper. Parameters varied are given in the caption of their respective figures. The number of sampling points varies with the FS length. We have a uniform spacing of $6\times10^{-3}/a$ between neighboring points. The value in Tab.~\ref{tab:Parameters} is a typical number.
\begin{table}
    \centering
    \caption{List of parameters}
    \begin{tabular}{c|c}
        \hline
        \textbf{Parameter} & \textbf{Value} \\ 
        \hline
        $a$ & $1$~\AA \\ 
        $\sigma_x$ & $0.4a$ \\ 
        $\sigma_y$ & $0.4a$ \\ 
        $t$ & $2.4$~eV\\
        $J_{\mathrm{sd}}S$ & $0.7$~eV\\
        $\eta$ & $-1.8$~N/m\\
        $\rho$ & $\sqrt{2}\eta$\\
        $\gamma$ & $2\eta$\\
        $M_\alpha$ & $12$~u\\
        $N_\mathrm{samp}$ & $800$\\
        \hline
    \end{tabular}
    \label{tab:Parameters}
\end{table}

\bibliography{Refs.bib}
\end{document}